\documentclass[onecolumn,epjc3]{svjour3}  
\smartqed  
\RequirePackage{graphicx}
\usepackage{acro}
\usepackage{subcaption}
\usepackage{color}
\renewcommand{\baselinestretch}{1.1}

\def\({\left(}
\def\){\right)}
\def\[{\left[}
\def\]{\right]}

\newcommand{\m}{{\rm m}}

\newcommand{\be}{{\begin{eqnarray}}}
\newcommand{\ee}{{\end{eqnarray}}}
\newcommand{\h}{{\rm km~s^{-1}Mpc^{-1}}}

\DeclareAcronym{LHAASO}{
  short = LHAASO ,
  long  = Large High Altitude Air Shower Observatory ,
  short-plural =  ,
}
\DeclareAcronym{GRB}{
  short = GRB ,
  long  = gamma-ray burst ,
  short-plural = s ,
}
\DeclareAcronym{CMB}{
  short = CMB ,
  long  = cosmic microwave background ,
  short-plural =  ,
}
\DeclareAcronym{EBL}{
  short = EBL ,
  long  = extragalactic background light ,
  short-plural =  ,
}
\DeclareAcronym{PDF}{
  short = PDF ,
  long  = probability distribution function ,
  short-plural = s ,
}

\RequirePackage[colorlinks,citecolor=blue,urlcolor=blue,linkcolor=blue]{hyperref}
\journalname{Eur. Phys. J. C}
\begin{document}

\title{Multi-TeV photons from GRB 221009A: uncertainty of optical depth considered }


\author{Zhi-Chao Zhao\thanksref{addr1}
        \and
        Yong Zhou\thanksref{addr2} 
        \and 
        Sai Wang\thanksref{e1,addr3,addr4} 
}

\thankstext{e1}{Correspondence author: wangsai@ihep.ac.cn} 


          \institute{Department of Applied Physics, College of Science, China Agricultural University, Qinghua East Road, Beijing 100083, People's Republic of China\label{addr1}
          \and 
          CAS Key Laboratory of Theoretical Physics, Institute of Theoretical Physics, Chinese Academy of Sciences, Beijing 100190, People's Republic of China \label{addr2}
          \and 
          Theoretical Physics Division, Institute of High Energy Physics, Chinese Academy of Sciences, Beijing 100049, People's Republic of China \label{addr3}
          \and 
          School of Physical Sciences, University of Chinese Academy of Sciences, Beijing 100049, People's Republic of China \label{addr4}
          }

\date{Received: date / Accepted: date}

\maketitle

\begin{abstract}
It is reported that the Large High Altitude Air Shower Observatory (LHAASO) observed thousands of very-high-energy photons up to $\sim$18 TeV from GRB 221009A. We study the survival rate of these photons via considering the fact that they are absorbed by the extragalactic background light. By performing a set of $10^6$ Monte-Carlo simulations, we explore the parameter space allowed by current observations and estimate the probability of predicting that LHAASO detects at least one photon of 18 TeV from GRB 221009A. We find that the standard physics is compatible with the observations of 18 TeV photons within $3.5\sigma$ confidence interval. Our research method can be straightforwardly generalized to study more data sets of LHAASO and other experiments in the future. 
\end{abstract}

\section{Introduction}
More than 5000 photons above 0.5 TeV emitted from GRB 221009A at redshift $z\simeq0.151$ \footnote{\url{https://gcn.gsfc.nasa.gov/gcn3/32648.gcn3}} were observed by the \ac{LHAASO} \footnote{\url{https://gcn.gsfc.nasa.gov/gcn3/32677.gcn3}} within 2000 seconds after the first detection by Swift, Fermi-GBM, Fermi-LAT, and so on \footnote{\url{https://gcn.gsfc.nasa.gov/gcn3_archive.html}}. The highest energetic photons were reported to reach $\sim18$ TeV, representing the first observation of photons above 10 TeV from \acp{GRB}. {\color{black} Such observations intrigued studies on photon mixing with axion-like particles \cite{Galanti:2022pbg,Galanti:2022xok,Baktash:2022gnf,Carenza:2022kjt,Gonzalez:2022opy,Lin:2022ocj,Troitsky:2022xso,Nakagawa:2022wwm,Zhang:2022zbm}, Lorentz symmetry violation \cite{Baktash:2022gnf,Li:2022vgq,Finke:2022swf,Zhu:2022usw,He:2022jdl,Huang:2022xto,Vardanyan:2022ujc,Li:2022wxc}, ultra-high-energy cosmic rays \cite{Das:2022gon}, dark photon \cite{Gonzalez:2022opy}, sterile neutrinos \cite{Cheung:2022luv,Smirnov:2022suv,Brdar:2022rhc}, and misidentification of the showers \cite{Baktash:2022gnf}}. In our work, we will investigate the survival probability of multi-TeV photons from GRB 221009A by considering the fact that they are significantly absorbed by the extragalactic background light intervening between the \ac{GRB} and the Earth. We will further show whether the standard physics is still capable to interpret current observations. 

Very-high-energy photons can dissipate their energies via annihilation with photons in \ac{CMB} and \ac{EBL}, producing electron-positron pairs. The threshold of this channel to happen is $E_{\mathrm{th}}=m_{e}^{2}/E_{\mathrm{b}}$, where $m_e$ and $E_{b}$ are the mass of electrons and the averaged energy of background light, respectively. Therefore, for \ac{CMB} photons, this threshold is hundreds of TeV, implying that we can safely disregard the effect of \ac{CMB} photons. However, the energy of \ac{EBL} photons can be higher by several orders of magnitude than the \ac{CMB} photons, changing the threshold to be lower by the same magnitude. For example, the threshold is $\sim2.6$ TeV if we consider $E_b\sim 0.1$eV. Therefore, we should take into account the suppression effect of \ac{EBL} photons on the detected flux of $\sim18$ TeV photons by \ac{LHAASO}. 

\section{Flux of TeV photons and EBL attenuation}
The \ac{EBL}-suppressed flux $F_{o}$ depends on the intrinsic flux $F_{i}$ and the optical depth $e^{-\tau}$ due to absorption by \ac{EBL}. Therefore, we have 
\begin{equation}
    F_{o}(E) = F_{i}(E) e^{-\tau(E,z)} \ ,
\end{equation}
where $E$ is the observed energy of photons and $z$ is the redshift of GRB 221009A. We use the tabulated data of the \ac{EBL} and optical depth measured by Ref.~\cite{Dominguez:2010bv}. The intrinsic flux of photons is approximated to be a power-law \cite{MAGIC:2019lau}  
\begin{equation}
    F_{i}(E) = A_{i} \left(\frac{E}{0.5\mathrm{TeV}}\right)^{\alpha_{\mathrm{i}}} e^{-\frac{E}{E_{\rm cut}}} \ ,
\end{equation}
where $A_{i}$ is an intrinsic flux at a pivot energy scale 0.5 TeV, $\alpha_{\mathrm{i}}$ denotes a spectral index, and $E_{\mathrm{cut}}$ is a cutoff energy scale. 
Based on the reports of Fermi-LAT, we have two measured values of $\alpha_i$, namely, $-1.87\pm0.04$ \footnote{\url{https://gcn.gsfc.nasa.gov/gcn3/32658.gcn3}} and $-2.12\pm0.11$ \footnote{\url{https://gcn.gsfc.nasa.gov/gcn3/32637.gcn3}}. However, they were obtained at 0.1--1 GeV, which is an energy range beyond the capability of \ac{LHAASO}. 
Meanwhile, there is not a report of spectral index from \ac{LHAASO} at present. 
During our parameter inference processes, we assume that the spectral index $\alpha_i$ is $-2$ and $-3$, respectively. 
Our results can be adjusted to fit any value of $\alpha_i$ between $-2$ and $-3$ if it is reported by \ac{LHAASO} in the future. 
Due to the same reason, we fix $E_{\mathrm{cut}}$ to 2 TeV, 5 TeV, and 10 TeV, respectively. 
Therefore, we leave $A_i$ to be determined by the data sets of \ac{LHAASO}.  

By considering the performance of \ac{LHAASO} \cite{LHAASO:2019qtb}, we predict the number of events within energy range $0.5-10$ TeV to be 
\begin{equation}
    N_{>0.5\mathrm{TeV}} = T \int_{0.5\mathrm{TeV}}^{10\mathrm{TeV}} F_{o}(E) S_{\mathrm{eff}}(E,\theta) dE \ ,
\end{equation}
where $S_{\mathrm{eff}}$ is the effective area of \ac{LHAASO}-WCDA, as provided in Fig.~6 of Chapter 3 in Ref.~\cite{LHAASO:2019qtb}, $\theta\simeq28^\circ$ is the zenith angle of GRB 221009A, and $T=2000$ seconds is the duration of \ac{LHAASO} observing run.

To explore the parameter space, we perform Bayesian analysis by considering the fact that the number of photons above 0.5TeV is more than 5000, as reported by \ac{LHAASO} \footnote{\url{https://gcn.gsfc.nasa.gov/gcn3/32677.gcn3}}. 
We assume that the event number follows Poisson distribution with \ac{PDF}, i.e.    
\begin{equation}
p(k)=\lambda^{k}e^{-\lambda}/k!
\end{equation}
with $\lambda=5000$ being the expectation and $k=N_{>0.5\mathrm{TeV}}$. Our results and conclusion are robust when choosing other value of $\lambda$ within a range $[5000,6000)$. If we consider smaller values of $\lambda$, e.g., $\lambda=4000$, our results would not be altered significantly, i.e., a difference of around 15\%, which would not destroy our conclusions. 
We implement the Bayesian inference by using the affine-invariant Markov chain Monte Carlo (MCMC) ensemble sampler in \textit{emcee} \cite{2013PASP..125..306F}. We assume that $A_{i}$ has a uniform prior in the range {\color{black}$[10^{-5},10^{-2.5}]$} in units of $\mathrm{TeV}^{-1}\mathrm{m}^{-2}\mathrm{s^{-1}}$. 
The optical depth is sampled by following the tabulated data of median value and uncertainties of $\tau$ \footnote{\url{http://side.iaa.es/EBL/}}, as described in Ref.~\cite{Dominguez:2010bv}.

\begin{figure}
\centering
\begin{minipage}{0.49\textwidth}
\centering
\includegraphics[width=\textwidth]{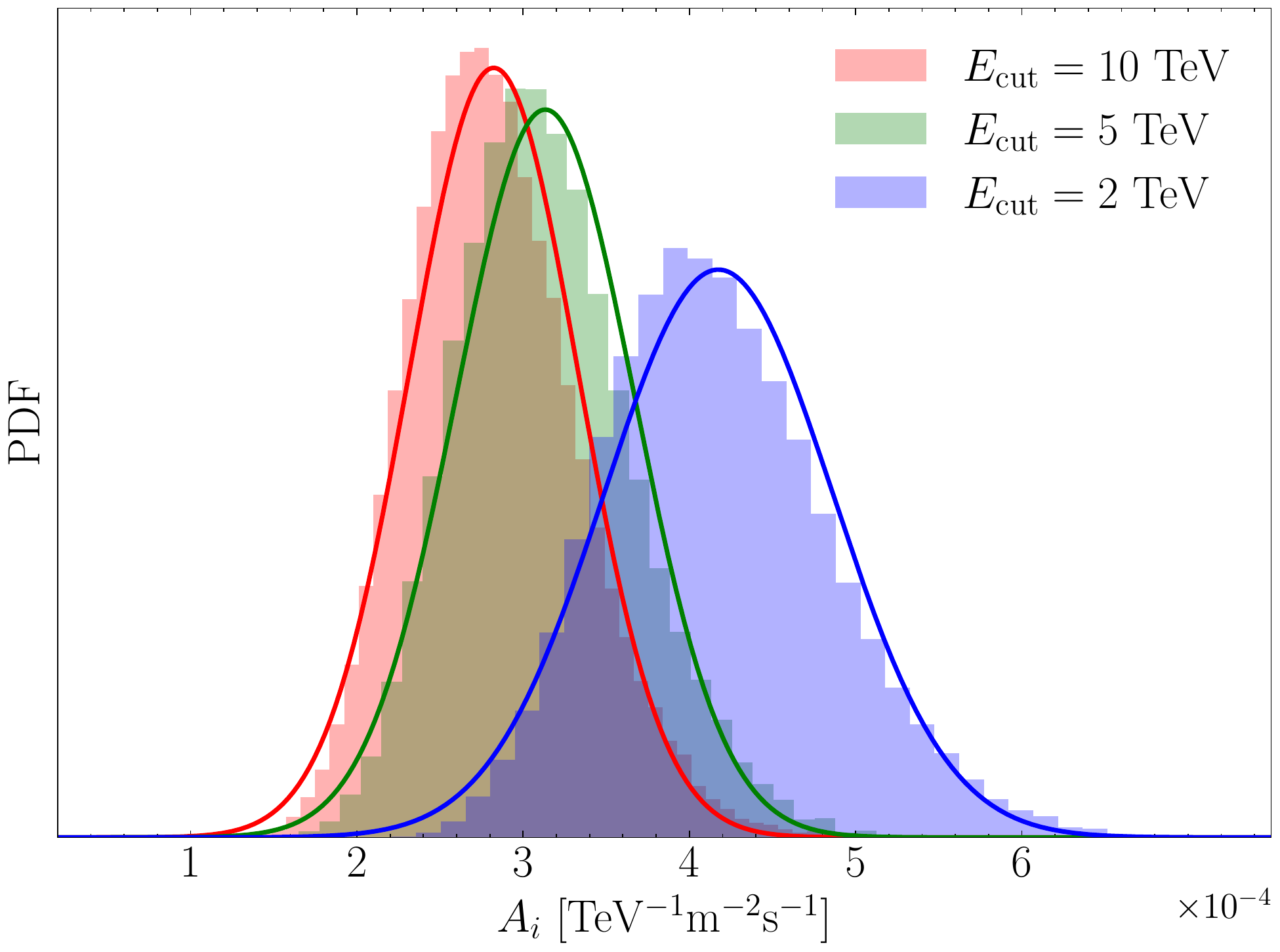}
\subcaption{$\alpha_i=-2$}
\end{minipage}\hspace{0.01\textwidth}
\begin{minipage}{0.49\textwidth}
\centering
\includegraphics[width=\textwidth]{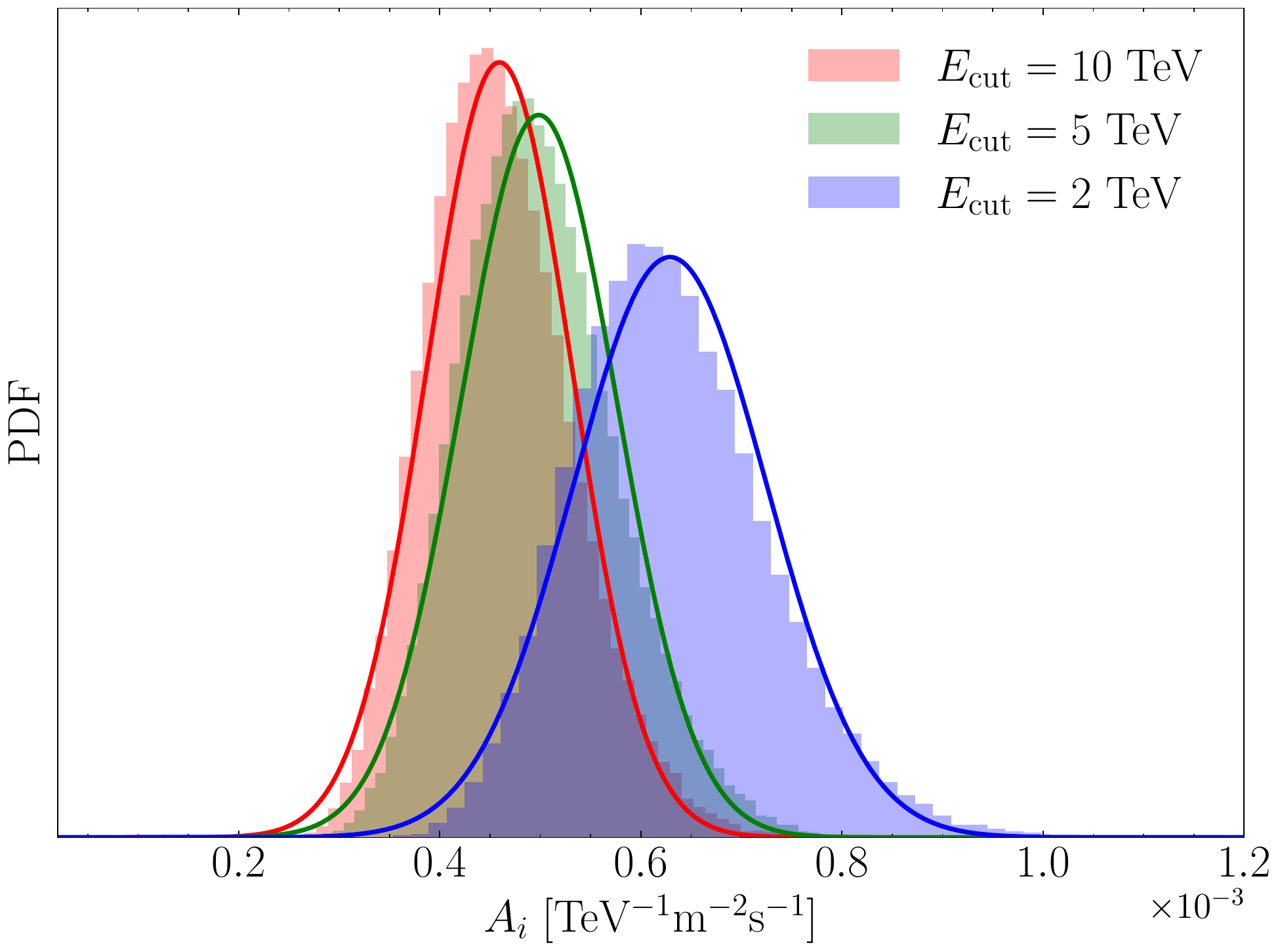}
\subcaption{$\alpha_i=-3$}
\end{minipage}
\caption{Posterior probability distribution functions of $A_{i}$ estimated in the case of $E_{\rm cut}= 10,\ 5,\ 2$ TeV for $\alpha_i=-2$ and $\alpha_i=-3$, respectively.}
\label{fig:figure1}
\end{figure}

The results of Bayesian parameter inferences are shown as 
the one-dimensional posterior \acp{PDF} of $A_i$ in Fig.~\ref{fig:figure1}. 
The left panel shows the results in the case of $\alpha_i=-2$ while the right one shows those in the case of $\alpha_i=-3$. 
For any case, we find $A_i\simeq\mathrm{few}\times10^{-4}$ $\mathrm{TeV}^{-1}\mathrm{m}^{-2}\mathrm{s^{-1}}$. 
In the following, we do Monte Carlo simulations via sampling $A_i$ following its posterior \ac{PDF}
and $\tau$ following the aforementioned tabulated data.

\section{Probability of detecting TeV photons}

Based on the above results, we will estimate the probability of predicting that \ac{LHAASO} observes at least one photon $\sim$18 TeV from GRB 221009A. During an observation of $T=2000$ seconds, the event number of photons with energy centered at $E$ is given by 
\begin{equation}
    N(E) = T \int_{10\mathrm{TeV}} F_{o}(E') S_{\mathrm{eff}}(E',\theta) P(E,E') dE' \ ,
\end{equation}
where $P(E,E') \propto \mathrm{exp}[-(E'/E-1)^{2}/(2\sigma^{2})]$ stands for a Gaussian \ac{PDF} with $\sigma=\Delta E/E$ being the energy resolution of \ac{LHAASO}-KM2A \cite{Cui:2014bda}.
We consider the energy range above 10 TeV, with an emphasis on 18 TeV. For each given energy $E$, we perform a set of $10^6$ Monte-Carlo simulations. We count the number of models that predict $N(E)\geq1$ and compute the corresponding probability via dividing this number by $10^6$.

\begin{figure}
\centering
\begin{minipage}{0.4\textwidth}
\centering
\includegraphics[width=\textwidth]{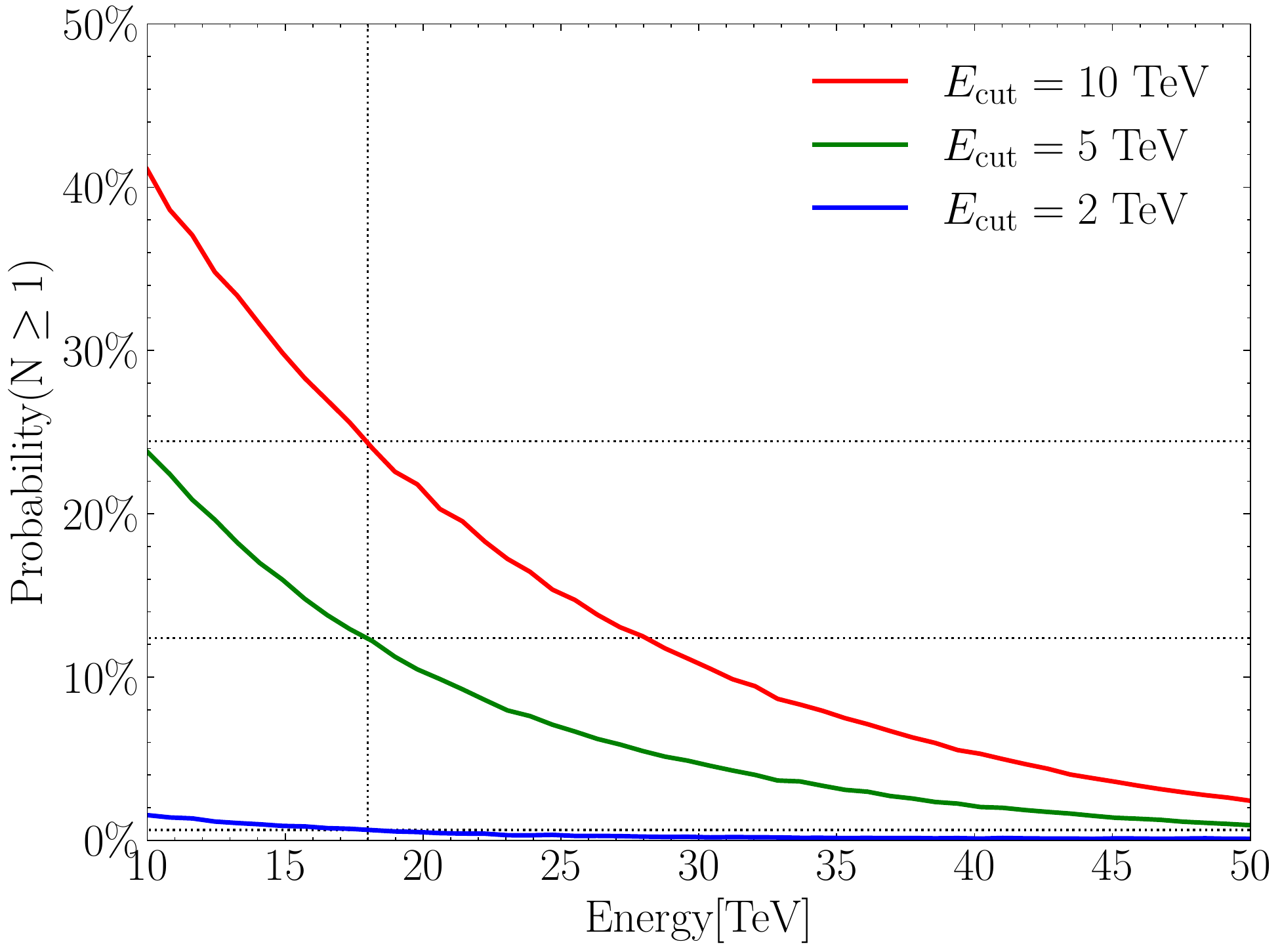}
\subcaption{$\alpha_i=-2$}
\end{minipage}\hspace{0.05\textwidth}
\begin{minipage}{0.4\textwidth}
\centering
\includegraphics[width=\textwidth]{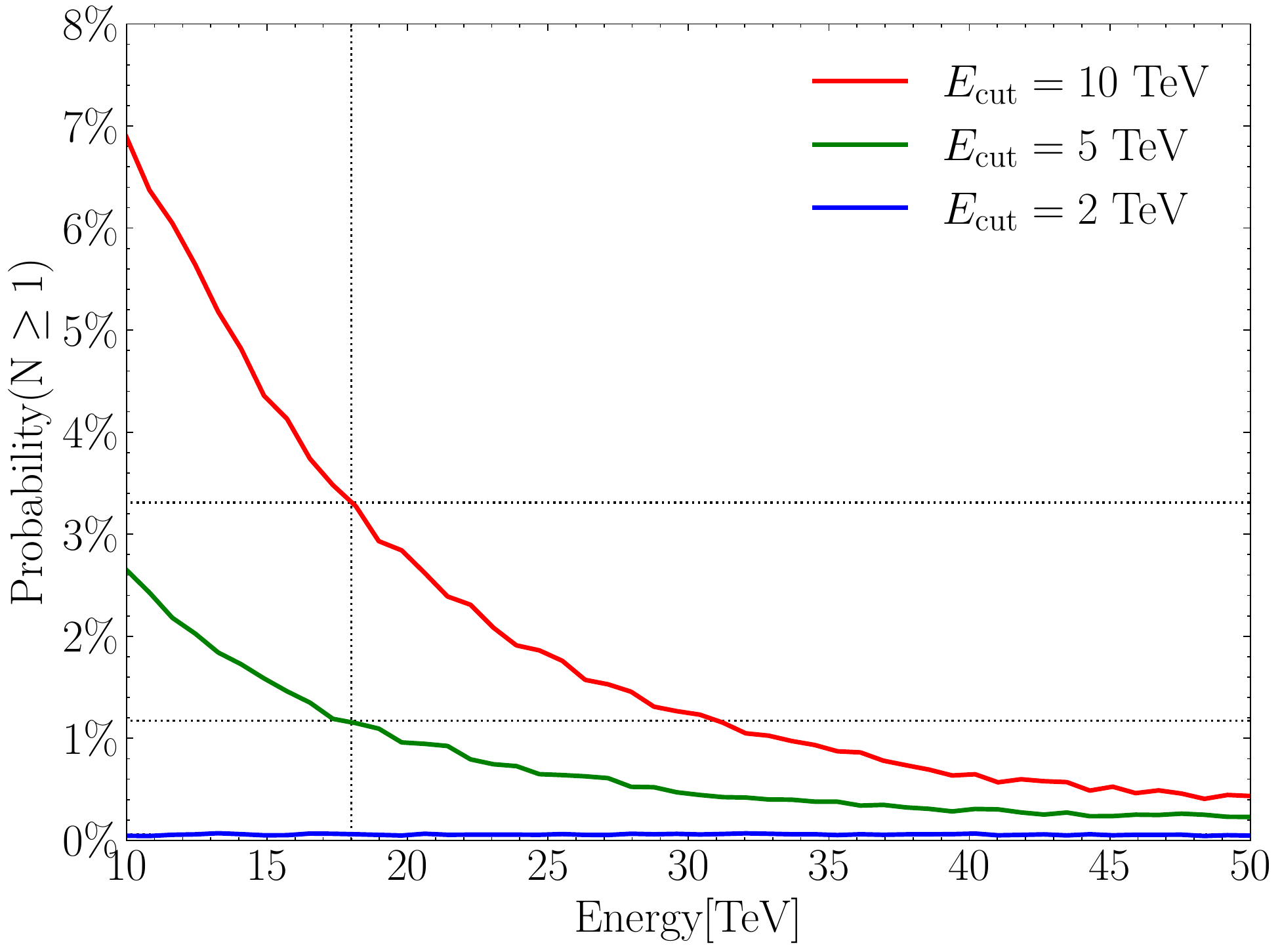}
\subcaption{$\alpha_i=-3$}
\end{minipage}
\caption{Probability of predicting that LHAASO observes at least one photon of multi-TeV from GRB 221009A within 2000 seconds for $E_{\rm cut}=10,\ 5,\ 2$ TeV respectively. The left panel shows the case of $\alpha_i=-2$ and the right panel shows the case of $\alpha_i=-3$. The dotted vertical lines denote 18 TeV photons while the dotted horizontal lines denote probabilities of observing at least one photon of around 18 TeV.}\label{fig:figure2}
\end{figure}

\begin{table}[]
    \centering
    \begin{tabular}{|c|c|c|c|}
    \hline
         --- & $E_{\rm cut} = 10 $ TeV  &   $E_{\rm cut} = 5 $ TeV &  $E_{\rm cut} = 2 $ TeV  \\
      \hline
       $\alpha_i=-2$  & 24.5 \% & 12.4 \% &0.6 \% \\
      \hline
       $\alpha_i=-3$  & 3.3 \% & 1.2 \% &  0.1 \%\\ 
      \hline
    \end{tabular}
    \caption{List for probabilities of predicting that LHAASO observes at least one photon of $\sim18$ TeV from GRB 221009A within 2000 seconds. }
    \label{tab:prob} 
\end{table}

Our results of Monte Carlo simulations are shown in Fig.~\ref{fig:figure2}. 
For a given energy, we estimate a probability of predicting that \ac{LHAASO} detects at least one photon. The left (right) panel shows the results in the case of $\alpha_i=-2$ ($\alpha_i=-3$). The labeling of $E_{\mathrm{cut}}$ is the same as in Fig.~\ref{fig:figure1}.  
In particular, the dotted vertical lines denote 18 TeV in the two panels, while the dotted horizontal lines denote the probabilities of predicting that \ac{LHAASO} is capable to detect at least one photon of 18 TeV from GRB 221009A. 
Correspondingly, we also list these probabilities in Tab.~\ref{tab:prob}. 
We find that in either case the standard physics is compatible with the observations of 18 TeV photons from GRB 221009A within $3.5\sigma$ confidence interval. 
This prediction could be further tested with the observational data sets of \ac{LHAASO} in the future.

\section{Summary}
In this work, we have investigated the survival rate of very-high-energy gamma rays within the energy range of \ac{LHAASO}, by taking into account the effect of \ac{EBL} attenuation. In the framework of standard physics, we simulated the probability of detecting the multi-TeV events from GRB 221009A. When considering the energy resolution of \ac{LHAASO}, we found that the standard physics is still compatible with the observations of 18 TeV photons from GRB 221009A within $3.5\sigma$ confidence interval. 
The above conclusions might be altered if we consider other measurements of \ac{EBL} and optical depth \cite{Saldana-Lopez:2020qzx,Gilmore:2011ks,Inoue:2012bk,Kneiske:2010pt,Finke:2009xi,Franceschini:2017iwq,Franceschini:2008tp}, that are beyond the scope of this paper. If the report of \ac{LHAASO} can be confirmed in the future, we may derive a novel constraint on the models of \ac{EBL} or even discriminate different models of \ac{EBL}. We would leave such detailed studies to future works. Our research method can also be straightforwardly generalized to study more data sets of \ac{LHAASO} and other experiments in the future.   
In addition, we did not consider other sources of astrophysical uncertainties, e.g., the effect of intergalactic magnetic fields. This effect may be considerable for GRB 221009A, especially in the Fermi-LAT energy band. In fact, the intergalactic magnetic field strength had been estimated to be $\sim10^{-16}$ Gauss, if considering the delayed cascade photons observed by Fermi-LAT, as shown in Ref.~\cite{Xia:2022uua}. 
Meanwhile, it was shown that \ac{LHAASO} might have observed the cascade photons from this GRB. However, it is challenging to discriminate them from photons due to other astrophysical processes in practice. 

We also noticed that other experiments detected photons from GRB 221009A, but at energy bands different from \ac{LHAASO}.  
For example, the Carpet-2 experiment \footnote{ATel $\#$15669} reported a single event $\sim251$ TeV in coincidence with GRB 221009A with statistical significance of $\sim3.8\sigma$. This event might imply an evidence of new physics due to such a high energy scale. However, another possibility was also proposed to be a cosmic-ray origin due to secondary emission from ultra-high-energy cosmic rays \cite{Das:2022gon,AlvesBatista:2022kpg,Mirabal:2022ipw}. 
Therefore, further observations in future are necessary to remove these debates.


\vspace{1em}
\begin{acknowledgements}
We acknowledge Prof. Xiao-Jun Bi, Prof. Hai-Nan Lin and Dr. Liang-Duan Liu for helpful discussions. 
ZCZ is supported by the National Natural Science Foundation of China (Grant NO. 12005016).
YZ is supported by the National Natural Science Foundation of China (Grants No. 12047558 and No. 12075249), the China Postdoctoral Science Foundation (Grant No. 2021M693238), and the Special Research Assistant Funding Project of CAS. 
SW is supported by the National Natural Science Foundation of China (Grant No. 12175243), the Key Research Program of the Chinese Academy of Sciences (Grant No. XDPB15) and the science research grants from the China Manned Space Project with No. CMS-CSST-2021-B01. 
\end{acknowledgements}

\bibliography{LHAASO-GRB221009A}
\bibliographystyle{h-physrev}

\end{document}